\documentclass[11pt]{article}

\usepackage{amsmath,amssymb}

\makeatletter

 \@addtoreset{equation}{section}
\makeatother

\newcommand {\beq}{\begin{eqnarray}}
\newcommand {\eeq}{\end{eqnarray}}
\newcommand {\non}{\nonumber\\}
\newcommand {\1}[1]{\frac{1}{#1}}

\newcommand {\ph}{\varphi}

\newcommand {\sig}{\sigma}

\newcommand {\dagg}{^{\dagger}}
\newcommand {\tr}{{\rm tr}\,}

\newcommand{\vs}[1]{\vspace{#1 mm}}
\newcommand{\hs}[1]{\hspace{#1 mm}}

\newcommand{\bsubeq}{\begin{subequations}}
\newcommand{\esubeq}{\end{subequations}}

\newcommand{\del}{\partial}

\begin{document}


\setcounter{page}{0}

\begin{titlepage}

{\normalsize
\begin{flushright}
{\tt hep-th/0411149} \\
TIT/HEP-532\\
\end{flushright}
}

\vs{2}

\begin{center}
{\huge\bf 
Conformal Sigma Models with Anomalous Dimensions
and Ricci Solitons}

\vs{30}
{\renewcommand{\thefootnote}{\fnsymbol{footnote}}
{\Large\bf 
Muneto Nitta\footnote{
     E-mail: {\tt nitta@th.phys.titech.ac.jp}}
}}

\vs{2}

{\it Department of Physics, Tokyo Institute of 
Technology \\
Tokyo 152-8551, JAPAN}

\end{center}

\setcounter{footnote}{0}


\vs{7}

\begin{abstract}

We present new non-Ricci-flat K\"ahler metrics 
with $U(N)$ and $O(N)$ isometries as target manifolds of 
superconformally invariant sigma models with 
an anomalous dimension.  
They are so-called Ricci solitons, 
special solutions to a Ricci-flow equation. 
These metrics explicitly contain the anomalous dimension 
and reduce to Ricci-flat K\"ahler metrics 
on the canonical line bundles over certain coset spaces 
in the limit of vanishing anomalous dimension.

\end{abstract}

\end{titlepage}

\section{Introduction}
String compactification with consistent 
non-trivial background 
is one of important subjects for long time.
Sigma model approach gives 
a set of equations of motion for such backgrounds 
by vanishing condition on 
the beta function~\cite{Polchinski:1998rq}.
Calabi-Yau manifolds, Ricci-flat K\"ahler manifolds, 
are required as compactified manifolds  
with constant dilaton backgrounds.
However no explicit metrics for 
compact Calabi-Yau manifold are known. 
On the other hand, if we allow non-compact manifolds, 
some explicit metrics for non-compact 
Calabi-Yau manifolds can be 
constructed~\cite{Ca}--\cite{Nitta:2003yk}.
Explicit solutions for non-Ricci-flat K\"ahler manifold
with a non-trivial dilaton 
background were obtained by 
Kiritsis, Kounnas and Lust \cite{Kiritsis:1993pb}
as a generalization of two dimensional 
Euclidean black hole~\cite{Wi,Hori:2001ax}. 

We can consider nonlinear sigma models whose 
scalar fields have an anomalous dimension 
instead of a constant dilaton background,  
because they can be transformed to each other 
by a field redefinition for the scalar fields 
to absorb the dilaton field~\cite{Hori:2001ax}. 
Using the Wilsonian renormalization group equation, 
the beta function of the ${\cal N}=(2,2)$ 
supersymmetric nonlinear sigma models consisting of 
the complex scalar fields $(\varphi^i,\varphi^{*i})$ 
with an anomalous dimension $\gamma$ was obtained in 
\cite{Higashijima:2002mh,Higashijima:2003ki,Higashijima:2003rp} 
(see \cite{Itou:2004px} for a review) 
as\footnote{
To derive this equation, 
it was useful to expand the Lagrangian in terms of 
K\"ahler normal coordinates~\cite{KNC} which 
are natural extension of Riemann normal coordinates 
to K\"ahler manifolds.
}
\beq
- {\del g_{ij^*} \over \del t} 
 = \beta(g_{i j^*})
 = \frac{1}{2\pi} R_{i j^*} 
  + \gamma \Big(\varphi^k g_{lj^*}\Gamma^l{}_{ik} 
  + \varphi^{* k} g_{i l^*}\Gamma^{l^*}{}_{j^*k^*} 
  + 2g_{i j^*} \Big) \label{CFT}
\eeq
with $g_{ij^*}$, $R_{i j^*}$ and $\Gamma^i{}_{ij}$ being  
the K\"ahler metric, the Ricci-form and 
the connection, respectively. 
Conformally invariant models are defined by 
the condition of vanishing beta function 
\beq
 R_{i j^*} + 2 \pi \gamma \Big(\varphi^k g_{lj^*}\Gamma^l{}_{ik} 
  + \varphi^{* k} g_{i l^*}\Gamma^{l^*}{}_{j^*k^*} 
  + 2g_{i j^*} \Big) 
  =0 .
   \label{RFeq}
\eeq 
A $U(N)$-invariant solution for this equation 
was obtained in \cite{Higashijima:2003ki}. 
It was shown to be equivalent to 
the $U(N)$-invariant solution with a dilaton background 
constructed in \cite{Kiritsis:1993pb}. 
This is due to the equivalence of this model to 
the dilaton model.
In this note we derive another $U(N)$-invariant solution 
and its extension to $O(N)$.

Eq.~(\ref{CFT}) is a so-called Ricci-flow equation 
which has attracted much attention recently in mathematics 
(see \cite{Bakas:2004hn} for a review by a physicist). 
In the Riemann manifold with a metric $g_{\mu\nu}$ 
and some vector field $\xi^{\mu}$, 
the general Ricci-flow equation is written as
\beq
 - {\del g_{\mu\nu} \over \del t}  
 = R_{\mu\nu} + \nabla_{\mu} \xi_{\nu} 
    + \nabla_{\nu} \xi_{\mu} . \label{Ricci-flow}
\eeq
If we take $\xi^{\mu} = 2\pi \gamma (\ph^i, \ph^{*i})$ 
as a special case in the K\"ahler manifold, 
the Ricci-flow equation (\ref{Ricci-flow}) 
reduces to Eq.~(\ref{CFT}).
Solutions to ${\del g_{\mu\nu} \over \del t} = 
R_{\mu\nu} + \nabla_{\mu} \xi_{\nu} 
    + \nabla_{\nu} \xi_{\mu}=0$ are called Ricci solitons 
and play a central role in classification of manifolds.
Our new solutions presented in this note 
are K\"ahler Ricci solitons and we hope that 
these solutions 
are useful for classification of K\"ahler manifolds.

\section{$U(N)$ Invariant Model}
We prepare an $N$-vector 
$\vec{\phi} = (\phi^1, \cdots, \phi^N)$ belonging 
to the fundamental representation of $U(N)$. 
Let us assume that the K\"ahler potential $K$ 
is written as a function of the $U(N)$-invariant as
\beq
  K = K(X) \,, \hs{5} 
  X \equiv \vec{\phi}\dagg \vec{\phi} .
\eeq
Geometric quantities can be calculated~\cite{Higashijima:2003ki}, 
to yield
\beq
 g_{ij^*} &=& K' \delta_{ij} + K'' \phi^{*i} \phi^j \\
 R_{i j^*} 
 &=& -\partial_i \partial_{j^*} \tr \log g_{k l^*} \nonumber\\
&=& -\left[(N-1)\frac{K''}{K'} +\frac{2K''+ K''' X}{K'+K'' X}  \right] 
         \delta_{ij} \nonumber\\
&& - \left[(N-1) \bigg(\frac{K'''}{K''} - \frac{(K'')^2}{(K')^2} \bigg) 
   + \frac{3K''' + K'''' X} {K' + K'' X} 
   - \frac{(2K'' + K''' X)^2}{(K' + K''X)^2} \right]
        \phi^{*i} \phi^{j}, \\
 g_{lj^*} \Gamma^l{}_{ik} &=& g_{ij^*},_{k} 
 = K'' (\phi^{*k} \delta_{ij} + \phi^{*i} \delta_{kj})
   + K''' \phi^{*k} \phi^{*i} \phi^j 
\eeq
with the prime denoting a differentiation with respect to $X$. 
Substituting these into Eq.~(\ref{RFeq}), 
we obtain an ordinary differential equation, 
from a term proportional to $\delta_{ij}$,
\beq
 (N-1)\frac{K''}{K'} +\frac{2K''+ K''' X}{K'+K'' X}
  + a (K' + K''X) = 0  \label{diffeq1}
\eeq
with $a$ a constant defined by the anomalous dimension 
as $a \equiv - 4 \pi \gamma$,  
and the derivative of this equation with respect to $X$ 
from a term proportional to $\phi^{*i} \phi^j$. 
The differential equation (\ref{diffeq1}) can be integrated to give
\beq
 (K')^{N-1} (K' + K''X) = c \, e^{- a K'X}  
  \label{eq-U}
\eeq
with $c$ an integration constant. 
Defining the function
\beq
 F \equiv K' X ,
\eeq
Eq.~(\ref{eq-U}) can be rewritten as
\beq
 F^{N-1} F' X^{1-N} = c \, e^{- a F} .
\eeq
Again this can be integrated to give the algebraic equation
\beq
  e^{a F}  
  \sum_{r=0}^{N-1} (-1)^r 
  {(N-1)!\, F^{(N-1)-r} \over (N-1-r)!\, a^{r+1} } 
   = {c \over N} X^N + b  \;  \label{eq-F}
\eeq
with $b$ an integration constant.
This reproduces the solution 
found in \cite{Kiritsis:1993pb,Higashijima:2003ki} with 
a boundary condition $F(0) = 0$ 
which implies $K(0)=$ const.   
It reduces to $F = \1{a} \log (1 + a X)$ for $N=1$ 
which defines the two dimensional 
Euclidean black hole~\cite{Wi,Hori:2001ax} 
for $a>0$.

We now construct a new $U(N)$-invariant solution.
To this end it is useful to define new coordinates by 
\beq
 \vec{\phi}^{\,T} = \sigma (1, \vec{z}^{\,T}) .
\eeq
We label $z$ by the same indices $i,j,\cdots$ with $\phi$
in the following. 
Then the invariant can be rewritten as
\beq
 X = \vec{\phi}\dagg \vec{\phi} 
  = |\sigma|^2 (1 + |\vec{z}|^2) 
   \equiv |\sig|^2 Z \, .
\eeq
It is useful to write down the metric in these coordinates as
\beq 
&& g = \left(
   \begin{array}{cc}
    g_{\sigma\sigma} & g_{\sigma j^*} \\
    g_{i \sigma^*}   & g_{ij^*}
     \end{array} 
    \right) , \non
&& g_{\sigma\sigma} = K'Z + K'' |\sigma|^2 Z^2 , \non
&& g_{\sigma j^*} = K' \sigma^* \del_{j^*} Z 
    + K'' \sigma^* |\sigma|^2 Z \del_{j^*} Z , \non
&& g_{ij^*} = K' |\sigma|^2 \del_i \del_{j^*} Z 
          + K'' |\sigma|^4 \del_i Z \del_{j^*} Z .
          \label{gen-metric}
\eeq
Using a solution $F$ of the same equation (\ref{eq-F}) 
with a boundary condition $F(0)= {\rm const.} \neq 0$ 
{\it different} from the previous one, 
the metric can be calculated 
in these coordinates as
\beq
&& ds^2 = 
    c \, e^{- a F} F^{1-N} Z^N |\sig|^{2N-2} |d \sig|^2
  + [c e^{- a F} F^{1-N} Z^{N-1} 
    |\sig|^{2N-2} \sig^* \del_{j^*} Z d \sig dz^{*j} 
      + {\rm c.c.}] \non
&& \hs{7} 
 + [ F (Z^{-1}\del_i \del_{j^*} Z - Z^{-2} \del_i Z \del_{j^*}Z)
    + c e^{- a F} F^{1-N} Z^{N-2} 
      |\sig|^{2N} \del_i Z \del_{j^*} Z] dz^i dz^{*j}. 
\eeq
Due to that boundary condition, 
this has a coordinate singularity in the limit $\sig \to 0$.
We perform the coordinate transformation 
\beq 
 \rho = \sigma^N / N   \label{coord}
\eeq
to remove this coordinate singularity.
We thus obtain the final form 
\beq
&& ds^2 = 
    c \, e^{- a F} F^{1-N} Z^N |d \rho|^2
  + [c N e^{- a F} F^{1-N} Z^{N-1} 
    \del_{j^*} Z \rho^* d \rho dz^{*j} + {\rm c.c.}] \non
&& \hs{7} 
 + [ F (Z^{-1}\del_i \del_{j^*} Z - Z^{-2} \del_i Z \del_{j^*}Z)
    + c N^2 e^{- a F} F^{1-N} Z^{N-2} 
      |\rho|^2 \del_i Z \del_{j^*} Z] dz^i dz^{*j}  
      \label{U-sol}
\eeq
with $Z = 1 + |\vec{z}|^2$.

The metric at the $\rho=0$ surface 
\beq
 ds^2|_{\rho =0} = F(0) 
  (Z^{-1} \del_i \del_{j^*} Z - Z^{-2} \del_i \del_{j^*}Z) dz^i dz^{*j}  
  =  F(0) \del_i \del_{j^*} \log (1 + |z|^2) dz^i dz^{*j}  
  \label{CPN}
\eeq
is the Fubini-Study metric on 
${\bf C}P^{N-1} \simeq SU(N)/[SU(N-1) \times U(1)]$. 
Therefore the metric (\ref{U-sol}) is 
the (canonical) line bundle over ${\bf C}P^{N-1}$, 
${\bf C} \ltimes {\bf C}P^{N-1}$. 
In fact, if we take $a=0$ we recover the Ricci-flat 
K\"ahler metric of Calabi~\cite{Ca} (see also \cite{HKN4}).

In the limit of 
the boundary condition constant $F(0)$ tending to zero, 
${\bf C}P^{N-1}$  with the metric (\ref{CPN}) shrinks and 
the whole metric (\ref{U-sol}) contains a singularity. 
It is an orbifold singularity in the orbifold 
${\bf C}^N/{\bf Z}_N$ defined by 
the identification (\ref{coord}).
Therefore $F(0)$ is a blow-up parameter 
for the orbifold singularity.

Since the asymptotic form of $F$ is 
$F \simeq {N\over a} \log aX$ for large $X$,
the metric becomes asymptotically 
\beq
&& ds^2 \simeq {c \over a X} (NX \log aX)^{1-N} Z^N |d \rho|^2
 + \left[ {c N \over a X} (NX \log aX )^{1-N} Z^{N-1} 
  \del_{j^*} Z \rho^* d\rho dz^{*j} + {\rm c.c.} \right] \non
&&\hs{7} 
 + \left[{N\over a} \log aX 
    (Z^{-1}\del_i \del_{j^*} Z - Z^{-2} \del_i Z \del_{j^*}Z) 
    + c N^2 \left({\log aX \over aX}\right)^{1-N} Z^{N-2} |\rho|^2 
     \del_i Z \del_{j^*} Z \right] dz^i dz^{*j} . \non
\eeq 

The difference between the previous solution 
\cite{Kiritsis:1993pb,Higashijima:2003ki} 
and the present solution is locally 
just the boundary condition. 
For the previous case, they required regularity on 
$K$ at $X=0$, $K(0)=$ const. (or $F(0)=0$), and therefore 
$K = k_0 + k_1 X + k_2 X^2 + \cdots$.
It is, however, not necessary for regularity on  
the metric as seen above. 
For our case the condition is $F(0) =$ const., and therefore 
$K = k_{-1} \log X + k_0 + k_1 X +   k_2 X^2 + \cdots$ 
is not regular at $X \to 0$. 
As a result the topology is drastically changed. 
The previous solution has topology ${\bf C}^N$, 
but the present solution has topology 
${\bf C} \ltimes {\bf C}P^{N-1}$ blowing up the orbifold singularity 
in ${\bf C}^N/{\bf Z}_N$. 

\section{$O(N)$ Invariant Model} 
Let us generalize the solution obtained in the last section 
to a $O(N)$-invariant solution.
We prepare an $N$-vector 
$\vec{\phi} = (\phi^A, \cdots, \phi^N)$ again  
and put a constraint 
\beq
 \vec{\phi}^{\,2} = \sum_{A=1}^N (\phi^A )^2 = 0 
  \label{constraint}
\eeq
to define a conifold~\cite{conifold}.
It is convenient to rewrite this constraint as
\begin{align}
 \vec{\phi}^{\, T} J \vec{\phi} = 0 , \hs{5}
J \ & \equiv \ \left(
\begin{array}{ccc}
0 & {\bf 0} & 1 \\
{\bf 0} & {\bf 1}_{N-2} & {\bf 0} \\
1 & {\bf 0} & 0 
\end{array} \right) 
\end{align}
with $J$ the rank-2 $O(N)$ invariant tensor.
The constraint (\ref{constraint}) 
can be solved as \cite{HN,Higashijima:2000rp}
\beq
 \vec{\phi}^{\, T} 
  = \sigma \left(1, z^i, -\1{2} \vec{z}^{\,2} \right)\; . 
   \label{QN-sol}
\eeq
Ricci-flat metrics on conifolds 
with the singularity deformed by $\vec{\phi}^{\,2} = r$
were constructed in \cite{conifold} for $N=3$ and 
\cite{Higashijima:2001yn} for general $N$.
Ricci-flat metrics on conifolds without deformation 
was constructed in \cite{HKN2,HKN4} 
which is still regular by an integration constant.
Our solution here is a non-Ricci-flat deformation 
of the latter one.

The $O(N)$-invariant can be written as 
\beq
 X \equiv \vec{\phi}\dagg\vec{\phi} 
  = |\sig|^2 \left( 1 + |\vec{z}|^2 + \1{4} |\vec{z}^{\,2}|^2 \right)
  \equiv |\sig|^2 Z . \label{Z-O}
\eeq
The expression of the metric is the same as the metric 
(\ref{gen-metric}) but with $X$ and $Z$ in Eq.~(\ref{Z-O}).
Components of the Ricci-form are calculated as
\beq
 && R_{\sig\sig^*} 
  = - L^{-1} L' Z - L^{-2} [L L'' - (L')^2] |\sig|^2 Z^2 ,\non
 && R_{\sig j^*} = 
  - L^{-1} L' \sig^* \del_{j^*} Z 
  - L^{-2}  [L L'' - (L')^2] \sig^* |\sig|^2 Z \del_{j^*} Z , \non
 && R_{ij^*} 
  = - L^{-1} L' |\sig|^2 \del_i \del_{j^*} Z 
   - L^{-2}  [L L'' - (L')^2] |\sig|^4 \del_i Z \del_{j^*} Z
\eeq
with $L \equiv (K')^{N-2} (K'' X^2 + K' X)$. 
Components of the connection are 
\beq
 && g_{\sig\sig^*,\sig^*} = K'' 3 \sig Z^2 + K''' \sig |\sig|^2 Z^3 ,\non
 && g_{\sig\sig^*,i^*} = K' \del_{i^*} Z 
  + K'' 3 |\sig|^2 Z \del_{i^*} Z + K'''|\sig|^4 Z^2 \del_{i^*}Z ,\non
 && g_{ij^*,\sig^*} = K' \sig \del_i \del_{j^*} Z 
    + K'' (2 \sig |\sig|^2 Z \del_i \del_{j^* }Z 
        + \sig |\sig|^2 \del_iZ \del_{j^*}Z) 
    + K''' \sig |\sig|^4 Z \del_i Z \del_{j^*} Z ,\non
&& g_{ij^*,k^*} = K' |\sig|^2 \del_i \del_{j^*} \del_{k^*} Z   
    + K''|\sig|^4 (\del_{k^* }Z \del_i \del_{j^* }Z 
        + \del_i Z \del_{j^*} \del_{k^* }Z 
        +  \del_{j^* }Z \del_i \del_{k^* }Z ) \non 
  && \hs{15} 
   + K''' |\sig|^6 \del_i Z \del_{j^*} Z  \del_{k^*} Z .
\eeq
Substituting all these quantities to Eq.~(\ref{RFeq}), 
we get 
\beq
 (\log L)' = - a (K' X)' .
\eeq
By integrating this we find that 
$F = K' X$ satisfies the equation similar to 
the $U(N)$ case,
\beq
 F^{N-2} F' X^{3-N} = c \, e^{- a F} 
\eeq
with an integration constant $c$.
Again this can be integrated to yield
\beq
  e^{a F} \sum_{r=0}^{N-2} (-1)^r 
  {(N-2)!\, F^{N-2-r} \over (N-2-r)!\, a^{r+1} } 
   = {c \over N-2} X^{N-2} + b  \;  \label{eq-F-O}
\eeq
with an integration constant $b$. 
The coordinate transformation 
\beq
 \rho = {\sigma^{N-2} \over N-2}
\eeq
is needed to remove the coordinate singularity. 
We thus obtain the final form of the metric 
\beq
&& ds^2 = 
    c \, e^{- a F} F^{2-N} Z^{N-2} |d \rho|^2
  + [c (N-2) e^{- a F} F^{2-N} Z^{N-3} 
     \del_{j^*} Z \rho^* d \rho dz^{*j} + {\rm c.c.}] \non
&& \hs{7} 
 + [ F (Z^{-1}\del_i \del_{j^*} Z - Z^{-2} \del_i Z \del_{j^*}Z)
    + c \, (N-2)^2  e^{- a F} F^{2-N} Z^{N-4} 
      |\rho|^2 \del_i Z \del_{j^*} Z] dz^i dz^{*j}  
      \label{O-sol}
\eeq
with $Z = 1 + |\vec{z}|^2 + \1{4} |\vec{z}^{\,2}|^2$. 

The metric at the $\rho=0$ surface is 
\beq
 ds^2|_{\rho =0} = F(0) 
  (Z^{-1} \del_i \del_{j^*} Z - Z^{-2} \del_i \del_{j^*}Z) dz^i dz^{*j}  
  =  F(0) \del_i \del_{j^*} 
    \log \left(1 + |z|^2 + \1{4} |\vec{z}^{\,2}|^2 \right) 
      dz^i dz^{*j}  .
  \label{QN}
\eeq
This is the metric on the quadric surface 
$Q^{N-2} \simeq SO(N)/[SO(N-2) \times U(1)]$~\cite{HN,Higashijima:2000rp}. 
The metric (\ref{O-sol}) is thus 
the (canonical) line bundle over $Q^{N-2}$. 
The Ricci-flat metric on a conifold \cite{HKN2,HKN4} 
is obtained by taking $a=0$ in the metric (\ref{O-sol}).
Our metric is non-Ricci-flat deformation of that conifold.

The asymptotic form of 
the metric (\ref{O-sol}) is ($F\simeq {N-2 \over a} \log aX$)  
\beq
&& ds^2 \simeq c [(N-2)X \log aX]^{2-N} Z^{N-2} |d \rho|^2 \non
&&\hs{7} 
   + \left[ c (N-2) \{ (N-2)X \log aX \}^{2-N} Z^{N-3} 
   \del_{j^*} Z \rho^* d\rho dz^{*j} + {\rm c.c.} \right] \non
&&\hs{7} 
 + \left[{N-2 \over a} \log aX 
    (Z^{-1}\del_i \del_{j^*} Z - Z^{-2} \del_i Z \del_{j^*}Z) \right. \non
&& \left. \hs{20}   
 + c (N-2)^2 \left\{ (N-2) X {\log aX}\right\}^{2-N} Z^{N-4} 
     |\rho|^2 \del_i Z \del_{j^*} Z \right] dz^i dz^{*j} . 
\eeq 

\section{Conclusion}
We have given two new metrics (\ref{U-sol}) and 
(\ref{O-sol}) with $U(N)$ and $O(N)$ symmetries, respectively  
as solutions of Eq.~(\ref{RFeq}) 
for target spaces of 
conformally invariant sigma models with 
an anomalous dimension $\gamma$. 
They are canonical line bundles over 
the projective space ${\bf C}P^{N-1}$ and 
the quadric surface $Q^{N-2}$, respectively.
These metrics contain the anomalous dimension 
explicitly through the parameter $a = - 4 \pi \gamma$. 
In the limit of vanishing anomalous dimension $a \to 0$, 
they reduce to those for Calabi-Yau manifolds~\cite{HKN2,HKN4}.
Generalization to other base coset spaces \cite{HKN4,HKN5} 
is straightforward. 
These new solutions give examples of K\"ahler Ricci solitons, 
singular solutions to the Ricci-flow equation (\ref{Ricci-flow}). 

A paper \cite{NS} recently posted to ArXiv has 
overlap with the present work. 
They have obtained explicit solutions 
with a dilaton background which also reduce to 
the same Calabi-Yau manifolds \cite{HKN2}--\cite{HKN6} 
with the present work.
Therefore solutions in the present paper and 
those in \cite{NS}
would be equivalent to each other 
by some field redefinition.

\section*{Acknowledgements}
The author would like to thank 
Kiyoshi Higashijima and Etsuko Itou 
for a collaboration in the early stages of this work 
and for explanation of their works 
\cite{Higashijima:2002mh,Higashijima:2003ki}. 
He is also grateful to Takashi Maeda for explanation of 
the Ricci-flow equation. 
His work is supported by Japan Society for the Promotion 
of Science under the Post-doctoral Research Program. 



\end{document}